\documentclass[fleqn,usenatbib]{mnras}

\usepackage{newtxtext,newtxmath}
\usepackage[T1]{fontenc}
\usepackage{ae,aecompl}

\usepackage{graphicx}	% Including figure files
\usepackage{amsmath}	% Advanced maths commands
\usepackage{amssymb}	% Extra maths symbols

\title[HaloNet: A deep CNN for dark matter haloes]{A volumetric deep Convolutional Neural Network for simulation of mock dark matter halo catalogues}

\author[P. Berger \& G. Stein]{
Philippe Berger,$^{1,3}$\thanks{E-mail: pberger@cita.utoronto.ca}
George Stein,$^{2,3}$\thanks{E-mail: gstein@cita.utoronto.ca}
\\
$^{1}$Department of Physics, University of Toronto,
 60 St. George St., Toronto, ON, M5S 1A7, Canada\\
$^{2}$Department of Astronomy \& Astrophysics,  University of Toronto, 50 St. George St., Toronto, ON, M5S 3H4, Canada\\
$^{3}$Canadian Institute for Theoretical Astrophysics,  University of Toronto, 60 St. George St., Toronto, ON, M5S 3H8, Canada\\
}

\date{Accepted XXX. Received YYY; in original form ZZZ}

\pubyear{2018}

\begin{document}
\label{firstpage}
\pagerange{\pageref{firstpage}--\pageref{lastpage}}
\maketitle

\begin{abstract}
For modern large-scale structure survey techniques it has become standard practice to test data analysis pipelines on large suites of mock simulations, a task which is currently prohibitively expensive for full N-body simulations. Instead of calculating this costly gravitational evolution, we have trained a three-dimensional deep Convolutional Neural Network (CNN) to identify dark matter protohaloes directly from the cosmological initial conditions. Training on halo catalogues from the Peak Patch semi-analytic code, we test various CNN architectures and find they generically achieve a Dice coefficient of $\sim\! 92\%$ in only 24 hours of training. We present a simple and fast geometric halo finding algorithm to extract haloes from this powerful pixel-wise binary classifier and find that the predicted catalogues match the mass function and power spectra of the ground truth simulations to within $\sim 10\%$. We investigate the effect of long-range tidal forces on an object-by-object basis and find that the network's predictions are consistent with the non-linear ellipsoidal collapse equations used explicitly by the Peak Patch algorithm.
\end{abstract}
%For modern large-scale structure survey techniques it has be become standard practice to test data analysis pipelines on large suites of mock simulations, a task which is currently prohibitively expensive for full N-body simulations. We have trained a three-dimensional deep Convolutional Neural Network (CNN) to identify dark matter proto-haloes from the cosmological initial conditions. A CNN operates directly on the initial density field allowing it to identify features that distinguish between halo and uncollapsed voxels. Training on halo catalogues from the Peak Patch semi-analytic code, we test various architecture variations and find they generically achieve a Dice coefficient of $\sim\! 92\%$ in only 24 hours of training. We present a simple and fast geometric halo finding algorithm to extract haloes from this powerful pixel-wise binary classifier and find that the predicted catalogues match the mass function and power spectra of the ground truth simulations to within $\sim 10\%$. We investigate the effect of long-range tidal forces on an object-by-object basis and find that the network's predictions are consistent with the ellipsoidal collapse equations used explicitly by the Peak Patch algorithm.
%\maketitle

%\tableofcontents

\section{\label{sec:intro} Introduction}

The fundamental observable in the study of the large-scale structure of the Universe is the non-linear matter density field. In N-body simulations of collisionless cold dark matter (CDM) particles, initially over-dense regions collapse under gravity to form virialized structures termed dark matter haloes. In the standard model of cosmology these objects form the potential wells in which baryonic matter can collect to form galaxies, galaxy groups, and galaxy clusters \citep{rotationcurves, fire}. An essential output of N-body simulations is a catalogue of positions, velocities, and masses of haloes. These mock dark matter halo catalogues allow us to interpret the observations of galaxy surveys and constrain cosmological models.

Modern large-scale structure survey techniques like Sunaeyev-Zeldovich effect \citep{plancksz,sptksz}, weak lensing \citep{planckwl,cfhtlens,kidswl,deswl}, or intensity mapping \citep{LIMreview}, hold incredible promise for constraining fundamental physics such as gravity on large scales \citep{kiyogr}, the equation-of-state of dark energy \citep{mmodes}, neutrino masses \citep{inman2016}, or the physics of inflation \citep{alvarezetal}. However each technique is accompanied by complicated systematics which, if not understood, would wash out the sought-after signal. It has become standard practice, therefore, to test data analysis pipelines on large suites of mock simulations \citep{avila2017,manera2013}, which combine realistic models of the signal and instrument. However, the number of simulations required to accurately determine survey error bars and scan parameter space is currently prohibitively large for full N-body simulations. This has led to the development of many `approximate methods' of large scale structure which attempt to create simulations of a satisfactory accuracy at minimal computational cost \citep{peakpatch,2013MNRAS.433.2389M,2013JCAP...06..036T,2016MNRAS.459.2327I,2016MNRAS.463.2273F,2015MNRAS.450.1856A,2014MNRAS.439L..21K,2015MNRAS.446.2621C,2014MNRAS.437.2594W}.  While full N-body simulations remain the most accurate tools available for modeling the dark matter of the Universe and mapping to observations, these approximate methods have been shown to be accurate at different spatial scales and levels of non-linearity, and are generally well suited for halo summary statistics and uncertainty quantifications.

In this work, we investigate the use of a Convolutional Neural Network (CNN) for fast generation of mock dark matter halo catalogues directly from the initial conditions. In recent years, CNNs have been lauded for their performance in computer vision tasks such as object detection or image segmentation \citep{nips2012}. CNNs have been shown to learn and identify features on multiple scales more efficiently than dense or fully-connected architectures, allowing to train deeper, more accurate models on larger datasets \citep{resnet}.   In cosmology, machine learning techniques have shown promise for the purposes of cosmological parameter estimation \citep{ravanbakhshetal,guptaetal,gilletetal}, simulating two-dimensional slices of the non-linear density field \citep{rodriguezetal}, initial conditions reconstruction \citep{modietal2018}, as well as classifying evolved structures in N-body simulations \citep{aragon-calvo2018} (who uses a similar CNN architecture to that of this work). Recently, \citeauthor{luciesmithetal} reported on their study of random forest classifier for haloes formed in an N-body simulation, traced back to the initial conditions.

Here we report the first application of a three-dimensional CNN for simulation of mock halo catalogs. We formulate halo-identification as a pixel-wise binary classification (or image segmentation) problem. The input of the CNN is therefore the initial, or Lagrangian space, density field and the output is a mask whose value is the network's certainty that a voxel ends up inside of a halo in the evolved simulation. The CNN is free to learn any spatial function (or feature) of the initial density field which allows it to distinguish between collapsed (halo) and uncollapsed voxels. This is unlike the random forest method of \cite{luciesmithetal}, where the input features are chosen.

The pioneering work of \cite{pressschechter} in the theoretical understanding of dark matter halo formation described the process statistically as a thresholding operation on the Gaussian random initial density field. \cite{bbks} added further constraints, noting that local maxima (or peaks) of the field should dictate the collapse dynamics, requiring information on both its first and second derivatives. \cite{peakpatch} formalized the relationship between tidal forces and tri-axial (or ellipsoidal) collapse of the regions surrounding peaks (peak patches), giving rise to the Peak Patch algorithm. The latter is the method used to generate the so-called ``ground truth'' simulations that we train our network on. The relationship between halo masses and tidal forces is both a non-trivial and a well defined property of (Peak Patch) haloes, which can be evaluated on an object-by-object basis. CNNs quickly learn to find edges \citep{nips2012}, for example, so it is interesting to ask whether more complicated combinations of derivatives can be learned as well.

In addition, producing large mocks of the universe with a CNN has two strong advantages over an N-body simulation: computational speed and orders of magnitude less memory requirement. The CNN by nature only considers the effects of pixels spaced by the size of the largest filter (which can be quite large, 128 Mpc in this study). This approximation allows the cosmological density field to be subdivided into separate volumes, eliminating the need to compute costly long-range gravitational forces, which typically requires expensive message passing between nodes. When combined with a multi-scale initial conditions generator (e.g. \cite{music}), holding the full simulation in memory at once can also be eliminated, and a large volume of the universe can be simulated on any modest machine.

The paper is outlined as follows. In the following section (\ref{sec:network}), we provide some relevant background information on the V-Net architecture we have adopted. Then, in Section \ref{sec:train}, we discuss the specifics of its implementation and the simulations that were used as training, validation, and test data. We tested several network variations in order to determine one that performed best after a reasonable amount of training. In Section \ref{sec:algo}, we suggest a simple and fast algorithm for extracting halo catalogues from the mask that is output by the CNN, with special attention to completeness. We call this method for producing mock catalogs ``HaloNet''. We can then evaluate the accuracy of these catalogues with population and clustering statistics, which we describe in Section \ref{sec:results}. Finally, in Section \ref{sec:discussion}, we discuss the implications of our findings and the future prospects for fast and accurate mocks with Convolutional Neural Networks.

\section{Convolutional Neural Networks for Image Segmentation}
\label{sec:network}

%BEGIN FIGURE -------------
\begin{figure*}
\begin{center}
\includegraphics[width=1.0\textwidth,trim={0 0 0 0},clip]{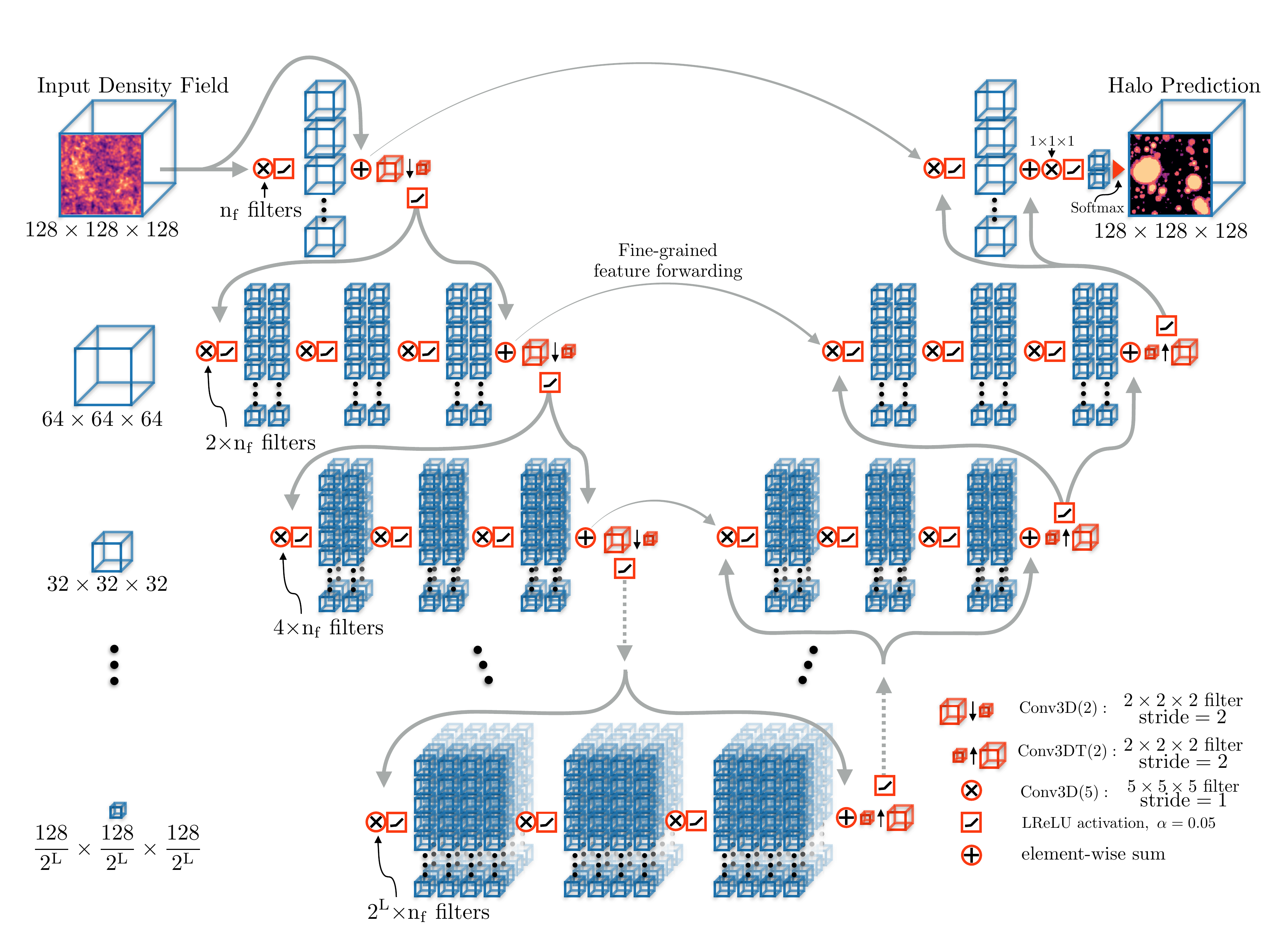}
\end{center}
\vspace{-0.3cm}
\caption{Schematic representation of our `V-net' architecture, where red symbols indicate all operations performed. The flow of data proceeds down the left side, identifying features on larger and larger scales of the 3-D input volume. These features are then remapped into the image space through de-convolutions as one re-ascends the right side. Through \textit{fine-grained feature forwarding} the features output from the same-dimension level on the left are concatenated onto those coming from below on the right. We show the general case of an \textit{L} level network, with an initial number of filters of $n_f$.} 
\label{fig:network}
\end{figure*}
%END FIGURE ----------------

A standard CNN passes inputs through a series of layers which decrease in size along the input dimension, but increase in size along a new dimension which labels the identified features. Eventually, at output, these features are combined to hopefully identify the image as belonging to one of the requested classes. The output dimension is equal to the number of classes and its value represents the network's certainty that input image is in that class. In image segmentation, however, the output should have the same dimensionality as the input, with values representing the certainty that an input element is part of a region of interest, termed the \textit{foreground}. This problem was addressed for medical image segmentation by \cite{unet}, whose U-Net architecture makes use of deconvolutions to return to the input image dimensionality. A deconvolution in this context is best understood as the transpose matrix operation of the standard convolution. We strongly recommend that the unfamiliar reader consult \cite{unet} and \cite{vnet} for detailed descriptions of the architectures. Figure~\ref{fig:network} shows a detailed schematic of the architecture used in our work and should be consulted before reading further. The term U-Net refers to a graphical picture of the U-shaped flow of the data through the network. One first descends the left side of the U, identifying features on larger and larger scales. These features are then remapped into the image space through deconvolution. \cite{unet} introduced the concept of \textit{fine-grained feature forwarding} where, as one re-ascends the right side of the U, the features output from the same-dimension level on the left are concatenated onto those coming from below. While the feature identification and deconvolution steps could in principle be trained separately, this has been seen to greatly speed up training by \citeauthor{unet}. The final convolutional layer's output has the same dimensionality as the input but with two features, across which a softmax $\sigma:\mathbb{R}^D \rightarrow [0, 1]^D$, is applied to convert the final output to a probability,
\begin{equation}
\sigma(z_i) = e^{z_i} / \sum\limits_{j=1,\dots,N}e^{z_j},
\end{equation}
where $i=1,\dots,N$, $D$ is the dimensionality of the space, and $N=2$. The two features are then compared to the ground-truth foreground and background masks, respectively, to compute the loss.

This idea was then applied in three-dimensions to MRI images by \citeauthor{vnet}, whose V-Net architecture we adopt for this work. \cite{vnet} further formulated the successive convolutions applied on each level (these don't change the dimensionality) as residual networks, which have been found to significantly improve the training of very deep networks \citep{resnet}.  We refer the reader to \cite{vnet} for further details on V-Net, however in Section \ref{sec:train} we summarize our implementation and the variations thereof that we have tested.

\section{Implementation and Training set}
\label{sec:train} 
\subsection{Peak Patch review}

We train our network on dark matter halo catalogues generated with the Peak Patch semi-analytic code,
described in \citep{stein2018,peakpatch,peakpatch2,peakpatch3}, and recently used to create large synthetic mocks of the extragalactic microwave sky\footnote{mocks.cita.utoronto.ca}, mocks of the carbon monoxide line emission from high redshift galaxies \citep{COMAP}, and to create covariance matrices of clustering statistics  \citep{Euclid1,Euclid2,Euclid3}.

Peak Patch identifies dark matter haloes in Lagrangian space by performing spherically averaged measurements of both the density and tidal tensor at locations of candidate peaks of the density field. A Peak Patch halo is therefore the largest spherical region of Lagrangian space which satisfies the conditions for ellipsoidal collapse at the target redshift. A hierarchical exclusion and binary merging algorithm is then performed to determine the final catalogues. A candidate peak is excluded if its centre lies inside a larger halo. If two peaks overlap only slightly, to conserve mass the overlapping mass is then subtracted from the smaller. The final haloes are then moved to their evolved positions (to Eulerian space) using second-order Lagrangian perturbation theory (although for this study we concentrate on the Lagrangian halo positions and masses, and so do not specify the details of the displacement). Peak Patch has passed extensive validations against many modern simulations, which will be outlined in a series of upcoming papers. 

For our training data, we have simulated 256 $(512~{\rm comoving\ Mpc})^3$ Peak Patch boxes, with 1 Mpc resolution (a 512$^3$ periodic grid). These are computed at redshift 0 and using the following cosmological parameters: $H_0=70.0$ km/s Mpc$^{-1}$, $\Omega_b=0.043$, $\Omega_c=0.207$, $\Omega_{\Lambda}=0.75$, $n_s=0.96$, and $\sigma_8=0.8$. We place a lower mass cutoff at the radius of a 27 cell halo. We keep 32 of these simulations as our testing set (Section \ref{sec:results}). The ground truth mask is generated at the same resolution from the Peak Patch catalogues by masking only voxels whose centres lie inside a halo. The masks and associated density fields are then divided into $128^3$ sub-volumes to be input to the network. Our training set therefore consists of 14336 independent volumes, an eighth of which are saved for validation. This is augmented by another factor of 8 by random reflections. While the Peak Patch algorithm itself requires both the density and displacement (velocity) fields, only the density is provided as input to the network.

We note that our method is also applicable to protohaloes traced back from the output of N-body simulations. While N-body simulations calculate the true dynamics of collisionless cold dark matter particles, dark matter haloes are dynamic objects with complicated morphologies \citep{diemerandkravtsov,adhikaridalalchamberlain}, and typical Eulerian space halo finders suffer from this uncertainty in their definition. We choose to perform this study on Peak Patch haloes largely for the ease with which we can generate large numbers of statistically-independent realizations, but also due to the unambiguous definition of a Peak Patch halo. In N-body, disconnected regions of Lagrangian space can belong to the same Eulerian halo (unlike in Peak Patch), which can add another degree of difficulty when performing Lagrangian spaced halo finding. Peak Patch has also been shown to have percent level accuracy for cluster and group sized haloes which are to a high degree spherical, and also reproduces a wide range of non-linear effects related to tidal forces also observed in N-body simulations. See Section \ref{sec:discussion} for a discussion of how our method generalizes. %Perhaps most importantly, for determining the accuracy of our network, we are not necessarily concerned with the degree to which our training set represents reality, but more how our network reproduces the training set. We leave the study of the V-Net architecture on N-body simulations and its accuracy for mock halo catalogs for later work. 

%BEGIN FIGURE -------------
\begin{figure}
\begin{center}
\includegraphics[width=1.0\columnwidth,trim={0 0 0 0},clip]{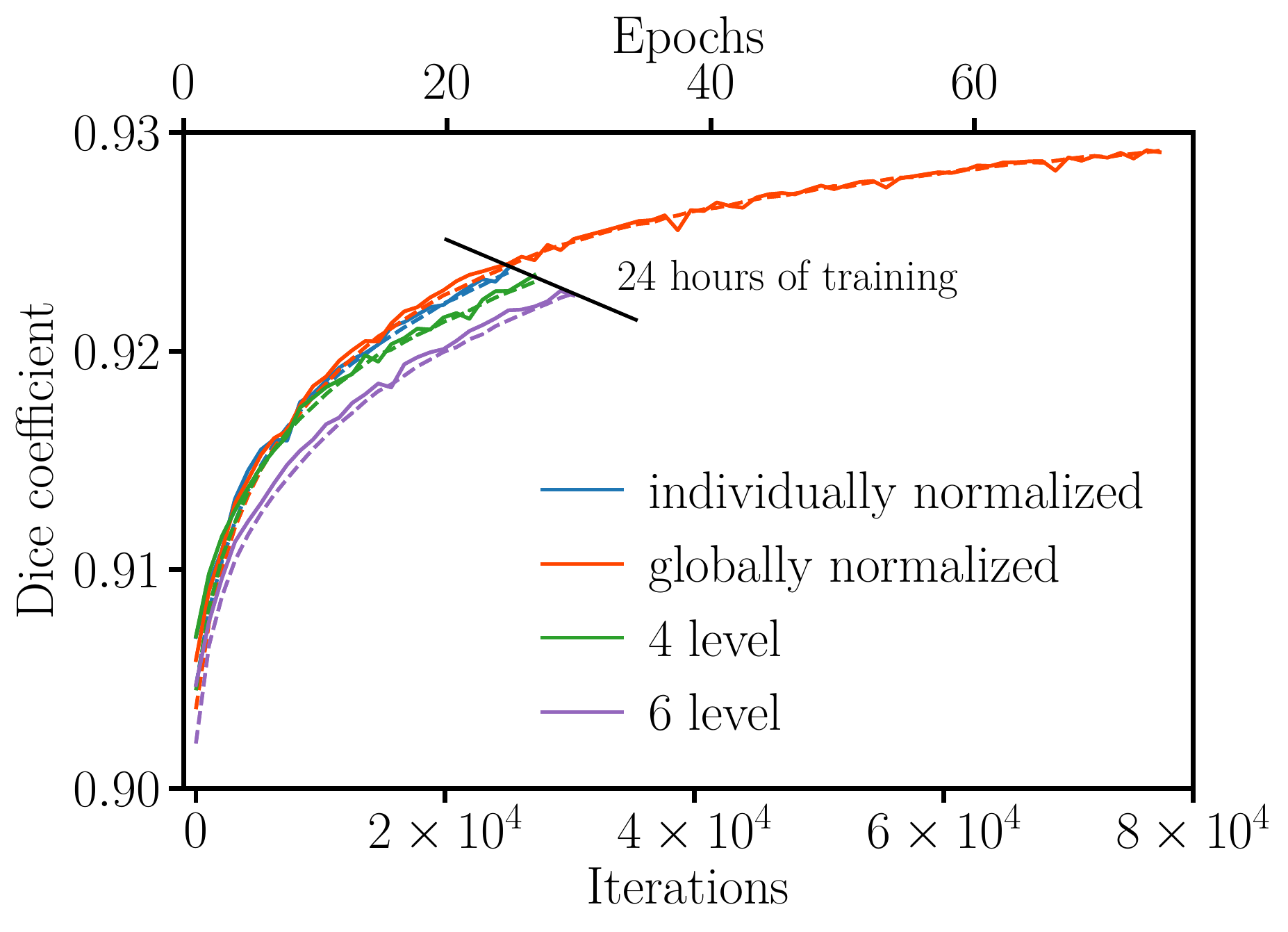}
\end{center}
\vspace{-0.3cm}
\caption{The Dice coefficient (Eq. \ref{dicecoeff}) is the loss function that we seek to maximize during training. The dashed and solid lines show the training and validation values respectively. We show the training curves for a 24 hour period for each of the architectures summarized in Table \ref{tab:architectures}. For this period, which occurs after an initial short stage of pre-training, we use a learning rate of 0.01, momentum of 0.9, and no dropout throughout (see text for further details). For the 5 level network, we show the curves with input training sets normalized by their grid-level standard deviation but also by the mean standard deviation of all simulations. The globally normalized 5 level is then trained for another 48 hours.}
\label{fig:train_plot}
\end{figure}
%END FIGURE ----------------

%------------ BEGIN TABLE -------------
\begin{table}
\centering
\caption{V-Net architectures tested}
\label{tab:architectures}
\begin{tabular}{llll}
\hline
Architecture & 4 level & 5 level & 6 level \\
\hline
\texttt{Conv3D(5)}/level & 3 & 3 & 3 \\ 
Initial filters $n_f$ & 16 & 16 & 10 \\
\texttt{Conv3D}\hyperlink{number1}{$^1$} & 29 & 36 & 43 \\ 
\texttt{Conv3DT}\hyperlink{number1}{$^1$} & 4 & 5 & 6 \\
LReLU\hyperlink{number1}{$^1$} & 32 & 40 & 48 \\
Free parameters & $2.38\times 10^8$ & $3.71\times 10^8$ & $4.00\times 10^8$ \\
\hline
\end{tabular}

\hypertarget{number1}{\hspace{-2.5cm} $^1$Total number in the entire network}
\end{table}
%------------- END TABLE --------------

\subsection{Implementation and training}
\label{sec:implementation}

%BEGIN FIGURE -------------
\begin{figure*}
\begin{center}
\includegraphics[width=0.9\textwidth,trim={0 0 0 0},clip]{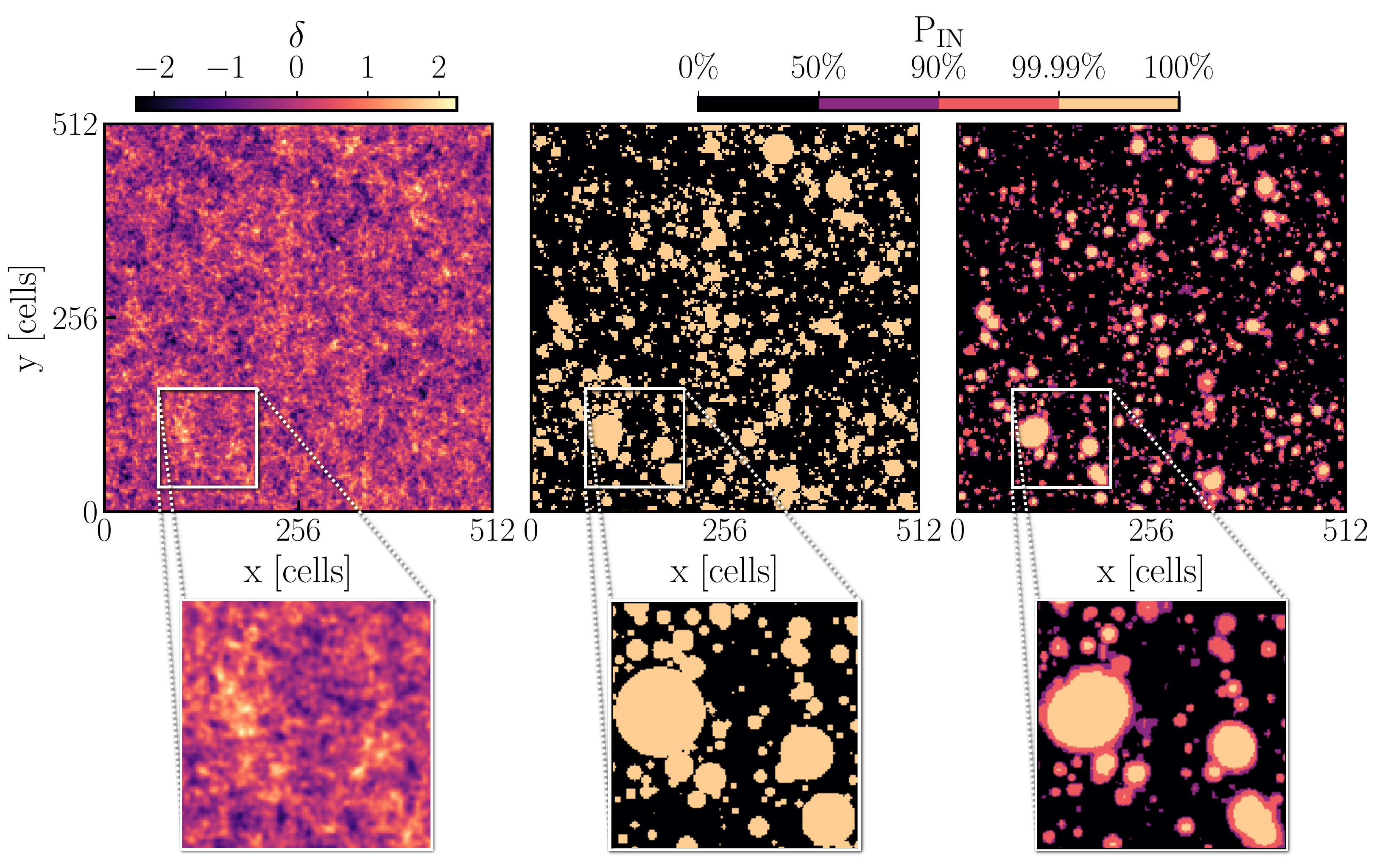}
\end{center}
\vspace{-0.3cm}
\caption{(left) A slice of the initial density field linearly extrapolated to redshift 0, where $\delta = \rho / \overline{\rho} - 1$. (middle) The collapsed regions of Lagrangian space belonging to haloes simulated using the Peak Patch method. (right) The HaloNet mask prediction, where the colour P$_{\rm{IN}}$ represents the certainty that a voxel belongs to a halo. Bottom panels show a zoom in on a high density region.} 
\label{fig:mask_visualization}
\end{figure*}
%END FIGURE ----------------

We have coded a custom implementation of V-Net using \texttt{keras} \citep{keras}. Following \cite{vnet}, we raise and lower the level (halve and double the resolution) with three-dimensional cubic down (\texttt{Conv3D(2)}) and up (\texttt{Conv3DT(2)}) convolutions of size 2 with stride of 2. On each level successive cubic convolutions of size 5 with unit stride are applied (\texttt{Conv3D(5)}), bracketed by an identity ``shortcut'' (see \cite{resnet} for details) to obtain a residual block. After every convolution we apply a leaky rectified linear unit (LReLU) activation with $\alpha=0.05$. Appropriate zero padding is used throughout to obtain the required output dimensionality. The architecture whose output we analyze in the following sections has 5 levels and 3 \texttt{Conv3D(5)}s within each residual block. This is true for all levels except the input, which performs a single \texttt{Conv3D(5)} to set the initial number of filters (in this case 16) and then a single \texttt{Conv3D(5)} bracketed by a shortcut. As our simulations have 1 Mpc resolution, this network down-samples to $2^5 = 32$ Mpc scales, but then learns $(5\times 5 \times 5)$ kernels on that level, meaning its largest filter could learn 160 Mpc features.

To test whether we are capturing large-scale environmental effects we train a 6 level V-Net as well. However, due to memory limitations we are forced to reduce the initial number of filters from 16 to 10 (which reduces the doubling on each successive level by that factor). Still, the total number of free parameters of the 5 and 6 level networks are comparable. To confirm that 5 levels are necessary to capture all information, we train a 4 level network with 16 initial filters. This information is summarized in Table \ref{tab:architectures}. All shallower networks, networks with a smaller number of filters, and networks with smaller convolution kernels performed worse in training than the 5 level. We also tested the use of batch normalization \citep{batchnorm} before non-linearities. While we observed gains in training smaller $64^3$ boxes with batch normalization at the input of every LReLU activation, the method has too large a memory overhead for the $128^3$ boxes. We tested several inhomogeneous placements of the normalizations but found that these were generally sensitive to overfitting.

We train using the stochastic gradient descent optimizer provided in \texttt{keras} and TensorFlow \citep{tensorflow} backend, on a single Power 8 node with 4 Nvidia Tesla P100 GPUs. Following \cite{vnet}, we maximize the Dice coefficient $\mathcal{D}$,
\begin{equation}
\mathcal{D} = \frac{2\vec{g}\cdot\vec{p}}{ |\vec{g}|^2 + |\vec{p}|^2},
\label{dicecoeff}
\end{equation}
where $\vec{g}$ and $\vec{p}$ are $D$-dimensional vectors representing the ground truth and network outputs. The sum is performed over both foreground and background masks. \citeauthor{vnet} proposed the Dice coefficient as a novel loss function for 2D image segmentation and found that it performed better than re-weighting methods for images with a strong background to foreground imbalance. While haloes are distributed differently than the foreground in that work, we find that training with the Dice coefficient proceeds quickly past the local minimum of an output volume filled with zeros.

Our training proceeds in two stages. For the first stage we use a learning rate of 0.1, momentum of 0.4, and dropout of 0.5 on the activations following the \texttt{Conv3D(2)} and \texttt{Conv3DT(2)} layers of the inner levels. We perform this stage a very small ($5 \times 512^3$ box) sample of the dataset. This allows the training to proceed very quickly past the local minimum corresponding the collapse fraction (the fraction of Lagrangian space which ends up in haloes above the minimum halo mass cutoff) of $f_{col} \simeq .27$ (i.e. Dice coefficient of $\sim 73\%$). For the second stage, we stop the training, change to a learning rate of 0.01, momentum of 0.9, turn off dropout, and iterate through the entire training set. These choices for the hyper-parameters were made by manually scanning the hyper-parameter space. While the choices we made yielded the fastest training, the network's ability to learn was largely robust to them. At all stages we use mini-batches of size 3, due to memory limitations.

%BEGIN FIGURE -------------
\begin{figure*}
\begin{center}
\includegraphics[width=1.\textwidth,trim={0 0 0 0},clip]{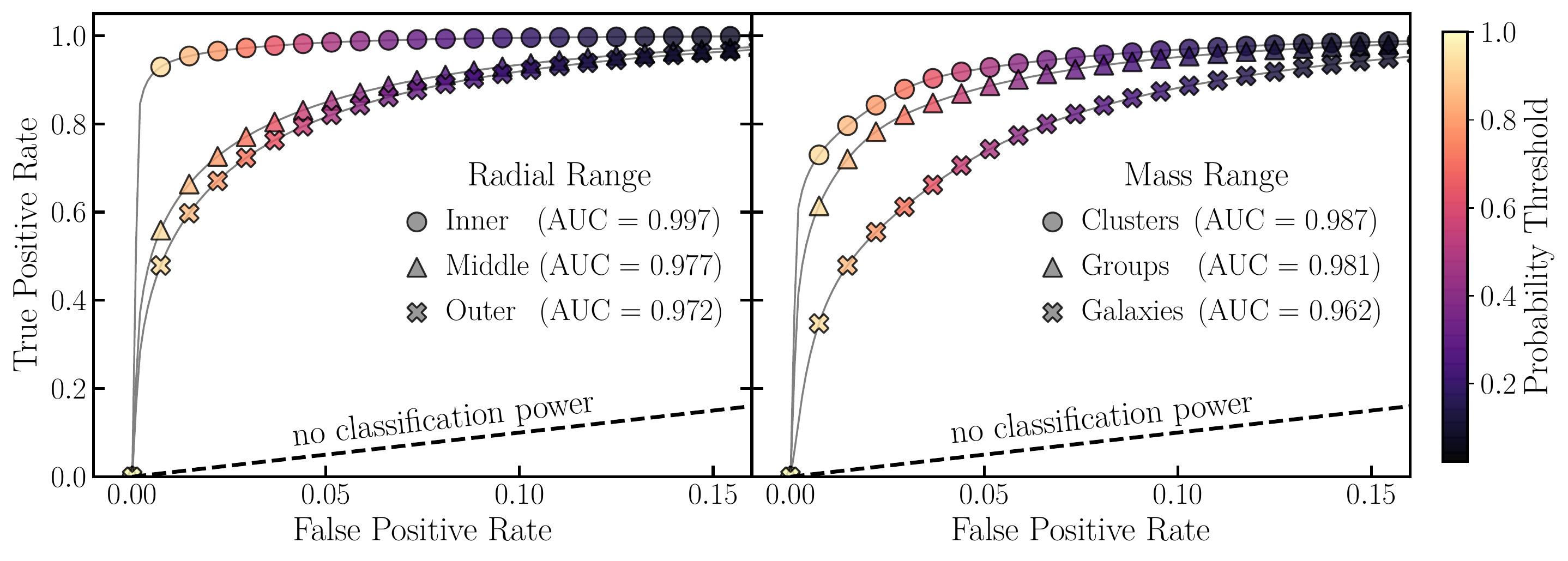}
\end{center}
\vspace{-0.3cm}
\caption{Receiver Operating Characteristic (ROC) curves to graphically represent the the true and false positive classification rates as a function of the probability threshold. See text for the definitions of the radial and mass ranges.}
\label{fig:ROC}
\end{figure*}
%END FIGURE ----------------

In Figure \ref{fig:train_plot}, we show the second stage of training for the architectures summarized in Table \ref{tab:architectures}. We train each for a 24 hour period, and find that all achieve a Dice coefficient $\geq 92 \%$. We find that the 5 level network achieves the largest Dice coefficient, despite the fact that its gradient updates (iterations) take the longest to compute, meaning the optimizer completes a smaller number of gradient updates in a fixed time. While the fraction-of-a-percent improvement of the 5 level over the 4 may seem marginal, as the Dice coefficient is a pixel-based quantity it is difficult to interpret (and the difference may be large) in terms of accuracy on halo properties. A better metric is to compare this increase only to the fraction of pixels in true haloes ($f_{col} \simeq .27$). Interestingly, we find a marginal improvement in normalizing each simulation by the mean grid-level standard deviation of all simulations versus its own, indicating that the 5 level network has access to information on box scales. Both the individually and globally normalized 5 level networks are shown in Figure \ref{fig:train_plot}, while the 4 and 6 level are individually normalized. We choose therefore to train the globally normalized 5 level for another 48 hours (72 hours total), and this final network in analyzed in the following sections. 

\subsection{Validating the Trained Neural Network}
\label{sec:train:subsec:train}

Having completed the training, we investigated the accuracy of the network's prediction for the set of 32 test simulations. HaloNet was trained on 128$^3$ volumes of the density field, so to predict the probability mask for full a 512$^3$ simulation (which represents the certainty that a voxel will end up in a dark matter halo) we partitioned it into sub-volumes of the input size and predicted on each separately. In order to avoid edge effects we split the simulation into 8$^3$ overlapping sub-volumes of 128$^3$ cells, where 32 cells on each edge are used as a buffer, as this is roughly the maximum halo size that we expect. Therefore, each pass to the network results in an effective volume of $(128-32\times2)^3 = 64^3$ cells. Combining the predicted sub-volumes back together allows us to create the final output mask. Due to the speed of HaloNet, the rather large buffer region (by relative volume) is not a computational problem, as a full 512$^3$ run takes only $\sim$3 minutes to predict.

In Figure~\ref{fig:mask_visualization} we see that the predicted mask visually resembles the true mask for the vast majority of haloes, which is not explicitly guaranteed by the high Dice coefficient. The panels in the center and right are coloured by the probability that a voxel, or ``dark matter particle'', ends up as part of a halo at redshift 0. Masked regions belonging to large haloes are correctly predicted to a very high level of accuracy. The predicted probability begins to decrease for smaller haloes, but even the smallest haloes almost all have a region of probability above the minimum cutoff shown of 50\%. This is promising for the halo finding we implement in the following section.

To characterize the performance of our pixel-wise binary classifier directly we create receiver operating characteristic (ROC) curves, which graphically represent the balance between the true and false positive rates as a function of the specified probability threshold. For a given probability cut, the true positive rate (TPR) and false positive rate (FPR) are given in terms of the number of true positives (TP), true negatives (TN), false positives (FP), and false negatives (FN), as
\begin{align}
TPR &= \frac{TP}{TP+FN},  \\
FPR &= \frac{FP}{FP+TN}.
\end{align}
True positives (negatives) in our classification correspond to particles correctly identified as ending up inside (outside) of haloes, while false positives (negatives) correspond to particles incorrectly identified as ending up inside (outside) of haloes, all for a given probability cut.

In Figure~\ref{fig:ROC}, we vary the probability threshold from 0 to 1 and calculate the TPR and FPR at each threshold in order to create a set of ROC curves. To quantify the performance of a classifier, a widely-used measure is the Area Under Curve (AUC). In the ideal case, the classifier would predict cells with 100\% accuracy at any threshold, and the AUC would be equal to 1. We separately calculated the ROC for mass ranges and radial ranges, adopting the same halo definitions of inner (r $<$ 0.3R$\rm _h$), middle (0.3R$\rm _h$ $<$ r $<$ 0.6R$\rm _h$), and outer (r $>$ 0.6R$\rm _h$) as \cite{luciesmithetal}, and similar definitions of clusters (M$\rm _h$ $>$ 10$^{14}$M$_\odot$), groups (10$^{13}$M$_\odot$ $>$ M$\rm _h $ $>$ 10$^{14}$M$_\odot$), and galaxies (3.2$\times$10$^{12}$M$_\odot$ $>$ M$\rm _h $ $>$ 10$^{13}$M$_\odot$). We find very high AUC values across all mass and radial splits, meaning HaloNet is a very accurate pixel-wise binary classifier for the problem in question. 

\begin{figure}
\begin{center}
\includegraphics[width=0.98\columnwidth,trim={0 0 0 0},clip]{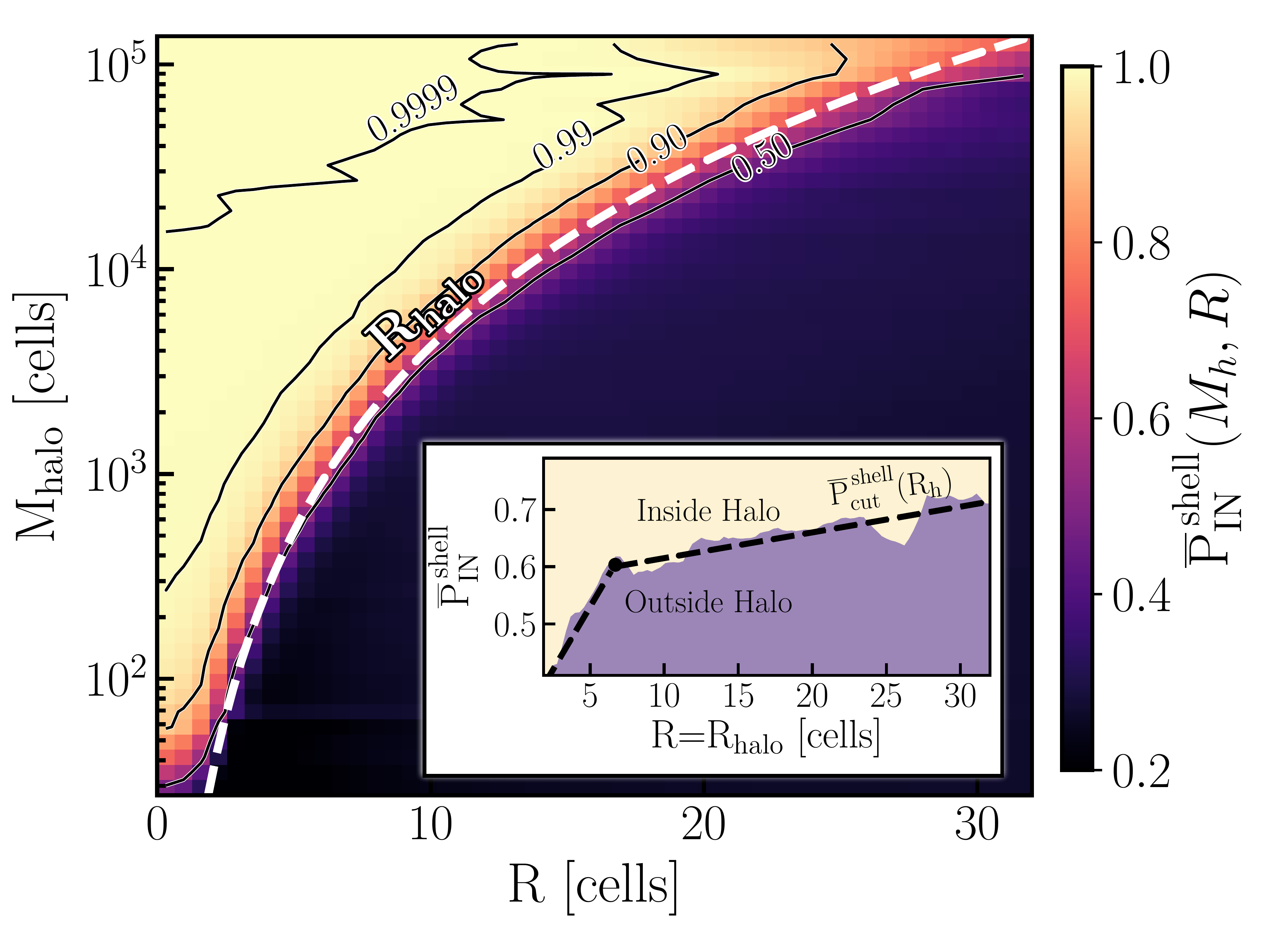}
\includegraphics[width=0.98\columnwidth,trim={0 0 0 0},clip]{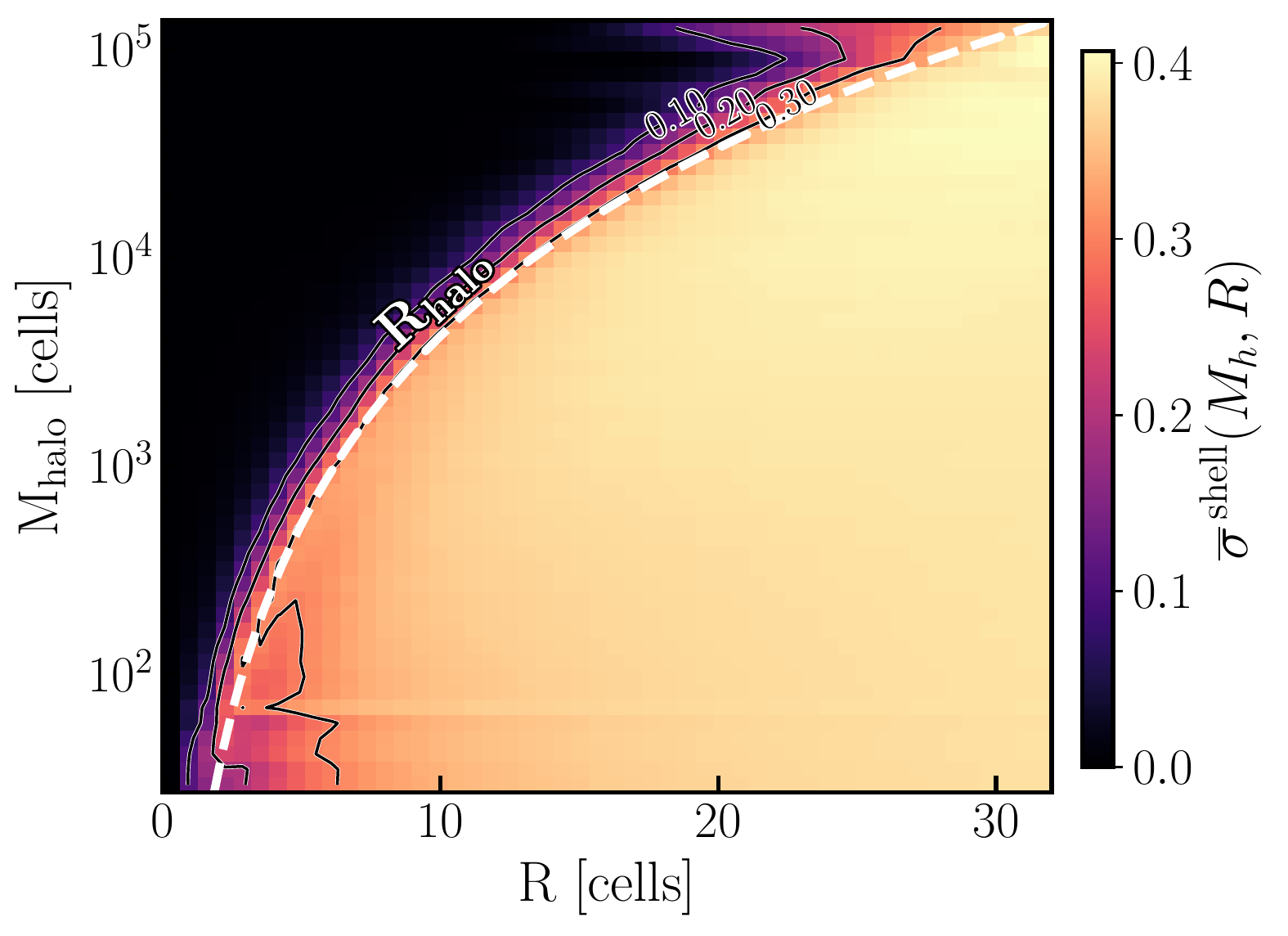}
\end{center}
\vspace{-0.3cm}
\caption{Calculating the spherical probability profile around the true locations of haloes allows us to address the accuracy of the mask directly. (top) The HaloNet predicted probability is averaged in shells as a function of distance from the center of true haloes. $\overline{\rm P}_{\rm{IN}}^{\,\rm{shell}}(M_h,R)$ is the average probability in a shell of radius R centered around true haloes of mass $M_h$. (top, inset) Shows the average probability values near the radius of haloes (the intersection of the white line). We find that a simple piecewise cut in probability as a function of radius matches very well with the size of the true halo, so we adopt this probability cutoff P$_{\rm cut}^{\rm shell}(R)$ to perform halo finding. (bottom) The standard deviation $\overline{\sigma}^{\,\rm{shell}}$ of the HaloNet mask stacked on the center of true haloes. The results shown are the average of 32 runs in order to decrease the noise from individual haloes, which is most apparent for the most massive haloes as these are the most rare.}
\label{fig:halo_measurements}
\end{figure}

The end goal of this work is to define haloes in the predicted probability mask to create a final halo catalogue. As we roughly want to maximize the TPR while minimizing the FPR, we could use the ROC to inform the choice of a constant probability cut to define as the boundary of haloes. But, as seen in the ROC curves, clusters are predicted with more accuracy than galaxies, providing motivation towards using an adaptive probability threshold as a function of scale. We therefore perform measurements of the probability profile around true halo locations to determine the average P$_\mathrm{cut}(\mathrm{R_{halo}})$ as a function of halo radius, seen in Figure~\ref{fig:halo_measurements}. The final probability thresholds we use typically lie in the range 0.5 $< P_\mathrm{cut}(\mathrm{R_{halo}}) <$ 0.7. From the ROC curves we see that this roughly corresponds to a false positive rate of 0.04 and a true positive rate of 0.8. We show in Section~\ref{sec:results} that these numbers are not directly related to the accuracy of the final halo catalogue, as halo finding can use local information and average the probability in radial shells to reduce many unwanted fluctuations in the probability field.   

%BEGIN FIGURE -------------
%END FIGURE ----------------

\section{Binary Classification to Halo Catalogue}
\label{sec:algo}

In order to transform a three dimensional mask of probabilities to a mock halo catalogue we need to partition volumes of the density field into individual dark matter haloes, based on the probability values of the predicted mask. In Figure~\ref{fig:mask_visualization} we can clearly see that predicted mask probabilities correspond with high accuracy to the true mask, but there remains some small differences, mostly due to overlapping haloes in high density regions. It is also apparent that regions of Lagrangian space belonging to more massive haloes have a greater central predicted probability in the HaloNet output when compared to smaller haloes. We use these observations to design a simple, hierarchical, geometrical, Lagrangian halo finder to identify haloes in the predicted mask, using three simple steps: 
\begin{enumerate}
\item Find connected regions in the field above a probability threshold P$_{\rm cut}^{\rm peak}$. The connected regions of space will be roughly non-spherical, but their centres-of-mass will nearly correspond to the centers of the the true haloes, given the predicted probability mask matches the true mask at those regions. The center-of-mass of each connected region is then used as the center of a new halo.
\item At each center, proceed outwards and average the probability mask in spherical shells until the mean probability of the shell has dropped below P$_{\rm cut}^{\rm shell}(\rm R_{\rm halo})$. To reduce the effects of halo clustering only consider cells that do not already belong to other haloes. The radius of the previous shell is then assigned as the radius of the halo, and the position and size of the halo are added to the final catalogue.

\item Descend to the next P$_{\rm peak}^{\rm cut}$ in the list P$_{\rm cut}^{\rm peak}=[p_0,p_1,...,p_n]$ and repeat steps 1-2. Using multiple probability thresholds of decreasing value ensures that small regions of the probability mask are not removed before the large haloes have been found.
\end{enumerate}

To determine P$_{\rm cut}^{\rm shell}(\rm R_{\rm halo})$ we stacked the HaloNet output on the true halo positions and measured the mean probability in radial shells outward from the origin, as seen in Figure~\ref{fig:halo_measurements}. We found that the radius of the true haloes corresponds roughly to $P_{\rm cut}^{\rm shell}(R_{\rm halo}) = 0.65$ for medium to large sized haloes, but the predicted probability begins to drop for smaller sized haloes. This is to be expected, as smaller haloes have larger tidal forces acting upon them, and are more difficult to predict. Therefore, we choose an empirically determined piecewise linear function, seen in the inset on the left of Figure~\ref{fig:halo_measurements}, which dictates approximately where the mean probability of a radial shell drops below P$_{\rm cut}^{\rm shell}$:
\[ P_{\rm cut}^{\rm shell}(R) =
\begin{cases} 
      0.044 x + 0.31 & R\leq 6.7~{\rm cells} \\
      0.0045 x + 0.57 & R> 6.7~{\rm cells}
   \end{cases}
\]
This set of radial probabilities is adopted as the definition of a HaloNet halo. We note that a flat probability cut of $P_{\rm cut}^{\rm shell} = 0.65$ gives similar results, but in order to maximize accuracy we adopted the piecewise linear function, as measuring it takes up a negligible amount of time compared to the training of the network.

%BEGIN FIGURE -------------
\begin{figure*}
\begin{center}
\includegraphics[width=.8\textwidth,trim={0 0 0 0},clip]{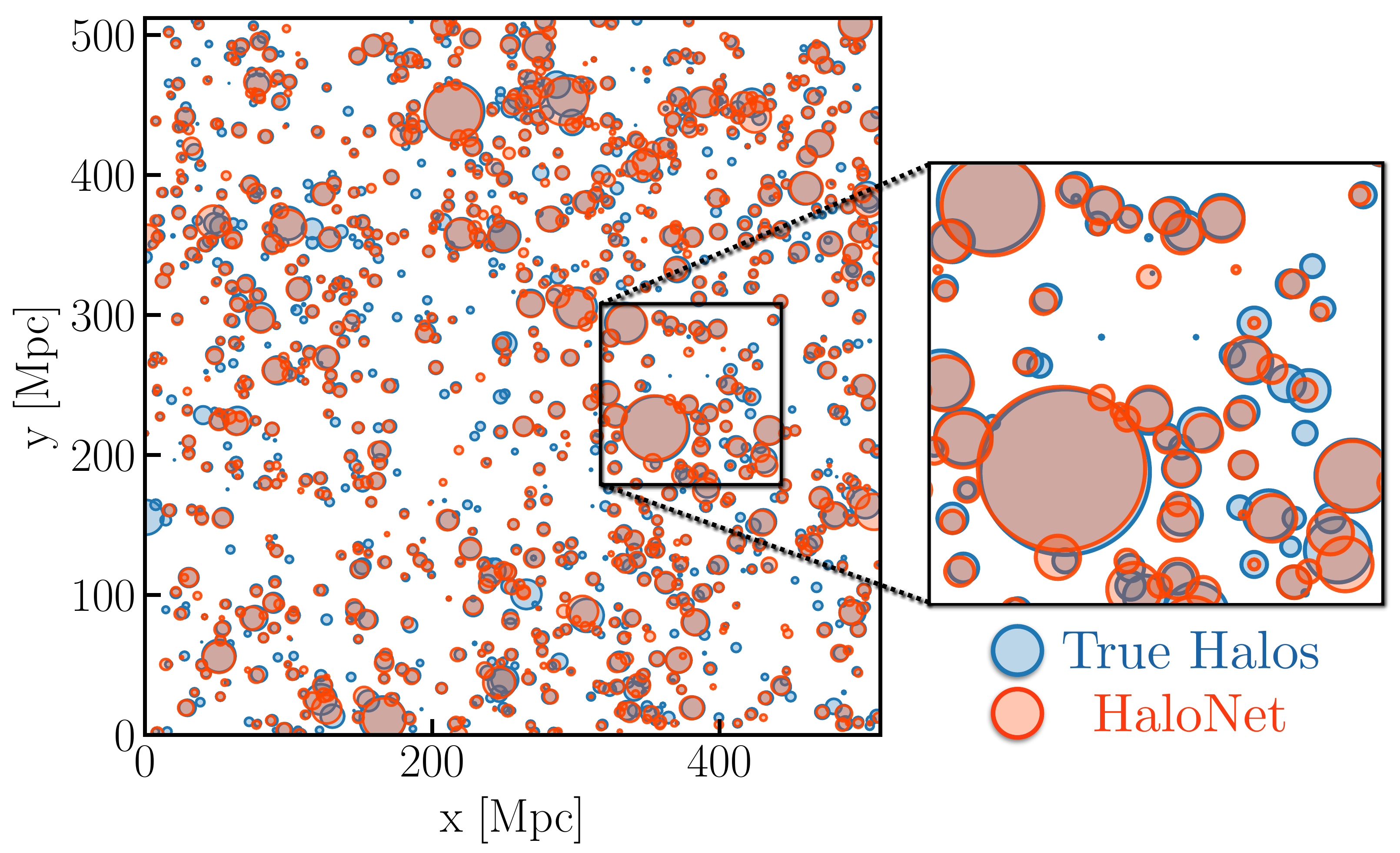}
\end{center}
\vspace{-0.3cm}
\caption{Results of our halo finder on the predicted probability mask compared to those of the true haloes. Shown here is a slice through the full simulation volume, where the size of the haloes plotted is their intersection with the slice. We find a near perfect prediction for large haloes, while some of the smallest haloes are less accurately found.}
\label{fig:halo_statistics}
\end{figure*}
%END FIGURE ----------------

In Figure~\ref{fig:halo_measurements} we also show the standard deviation in radial shells outward from the true halo centers. The standard deviation remains very close to 0 at small radii, meaning that the probability mask is very close to spherically symmetric around the locations of true haloes. Closer to the radius of the halo the standard deviation of the shell starts to increase, meaning that the mask starts to become less spherical near the halo boundary. This increase is largely due to halo clustering, so the probability mask is spherical to a good approximation within the radius of the halo.

The values of P$_{\rm cut}^{\rm peak}$ were determined by the probability contours of Figure~\ref{fig:halo_measurements}. By choosing a first P$_{\rm cut}^{\rm peak}$ of 0.9999 we will find all haloes above a mass of $2\times 10^4$ cells. As haloes of this mass are rare, these should be non-overlapping. Similarly, by choosing a second P$_{\rm cut}^{\rm peak}$ of 0.99 we will find all haloes above a mass of $3\times 10^2$ cells. We find that P$_{\rm cut}^{\rm peak}=[0.9999,0.99,0.975]$ results in an accurate halo catalogue, finding the required haloes across the whole mass range. We performed a simple convergence test, adding 3 extra P$_{\rm cut}^{\rm peak}$ values between each of the ones listed above, and adding filters below, and found no improvement. 

\section{Mock Catalogue Results}
\label{sec:results} 

Satisfied that our halo finder provides an accurate identification of the objects in the probability mask, we validate the HaloNet halo catalogues against the true halo catalogues using four main categories: visually, halo abundance, halo clustering, and halo measurements. For this study, we evaluate our mock catalogs in Lagrangian space since this is the minimally processed output of the Neural Network. A well established and computationally negligible method for moving halos to their final positions is second-order Lagrangian perturbation theory (2LPT). A large body of literature exists on 2LPT \citep{2LPT}.

In Figure~\ref{fig:halo_statistics} we show a comparison of final halo catalogues. It is immediately apparent that HaloNet accurately reproduces the mass, position, and clustering of the dark matter haloes.

\subsection{Halo Masses}
\label{sec:massfunction}

Correctly determining the abundance of haloes of a given mass is crucial when creating a mock catalogue. Many observables, such as the Sunyaev-Zel'dovich effect \citep{plancksz,sptksz}, weak lensing \citep{planckwl,cfhtlens,kidswl,deswl}, or line intensity \citep{LIMreview} can be directly related to the total mass and redshift of the cluster. Therefore, incorrect masses will inhibit the mock's ability to reproduce the statistics of true cosmological observations. 

The mass of a Peak Patch halo is defined by the largest spherical region which collapses by the redshift of interest under the homogeneous ellipsoid collapse approximation. As the periodic grid of the simulation is defined in comoving coordinates, the radius of a halo is easily related to the mass through $\rm{M_{h} = \frac{4}{3} \pi \rm R_{h}^3 \bar\rho_{M}}$, where $\rm{\bar\rho_M = 2.775 \times 10^{11} \Omega_M h^2\ [M_\odot/Mpc^3]}$ is the mean matter density of the universe. We adopt the same definition of mass for our HaloNet haloes.  

%BEGIN FIGURE -------------
\begin{figure}
\begin{center}
\includegraphics[width=.98\columnwidth,trim={0 0 0 0},clip]{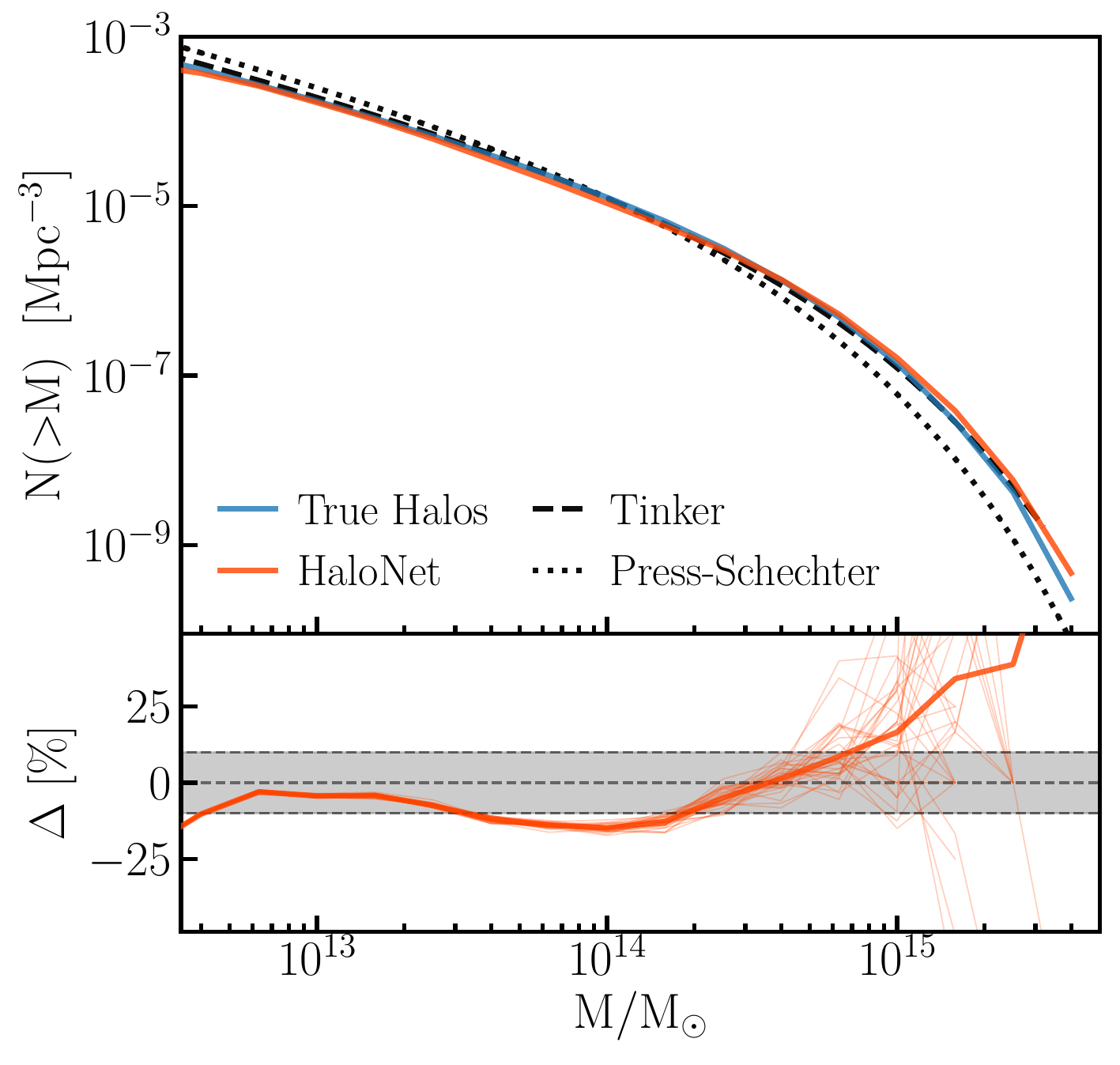}
\end{center}
\vspace{-0.3cm}
\caption{The mean halo mass function of the 32 test simulations. $\rm{N(>M)}$ is the number of haloes with a mass greater than M. The bottom panel shows the difference of the HaloNet massfunction compared to the massfunction of the true haloes. The thick line is the mean, while the thin lines are each of the test simulations individually. The shaded grey area represents a 10\% deviation.}
\label{fig:halo_massfunction}
\end{figure}
%END FIGURE ----------------
%BEGIN FIGURE -------------
\begin{figure}
\begin{center}
\includegraphics[width=0.48\textwidth,trim={0 5 0 0},clip]{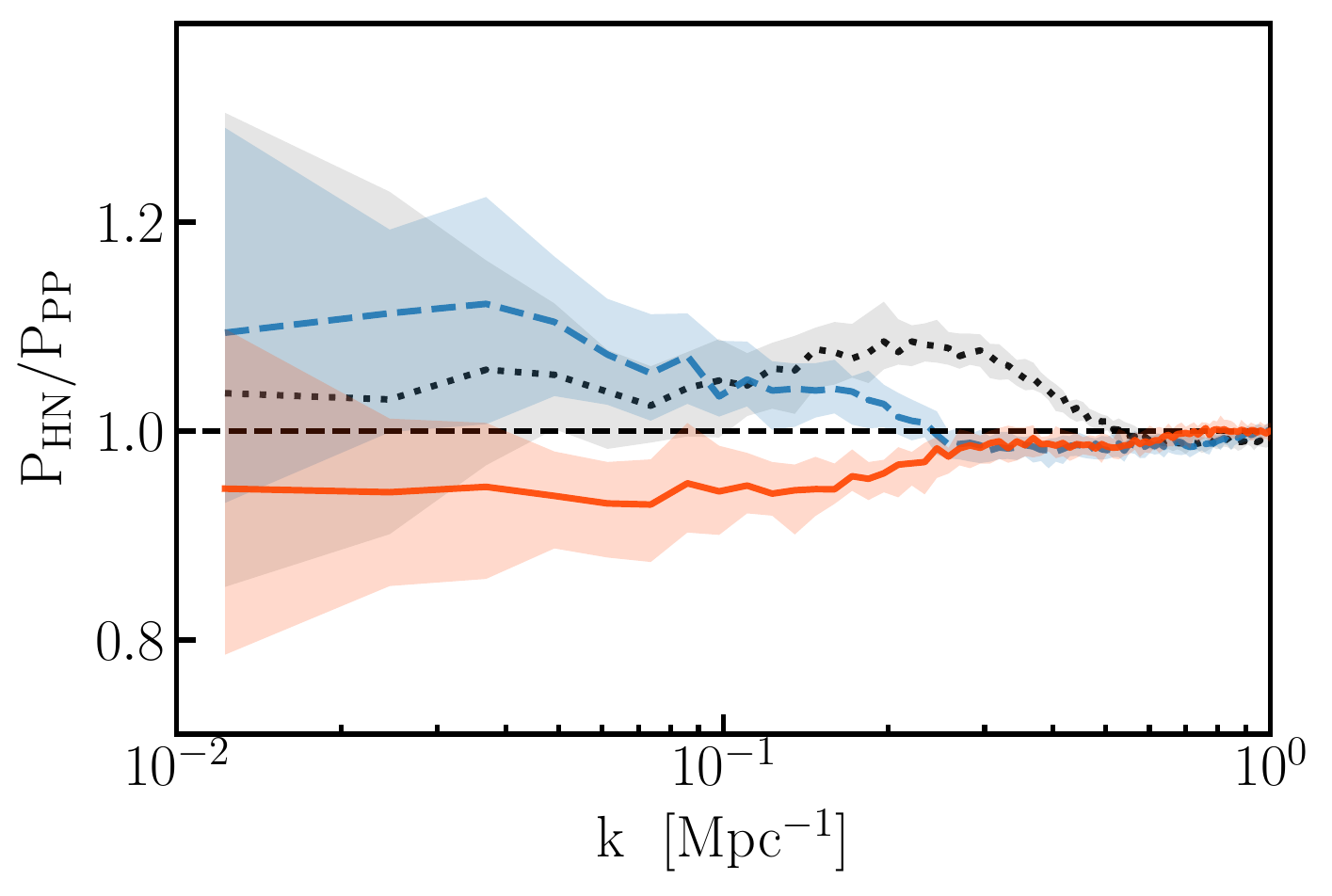}
\includegraphics[width=0.48\textwidth,trim={0 5 0 5},clip]{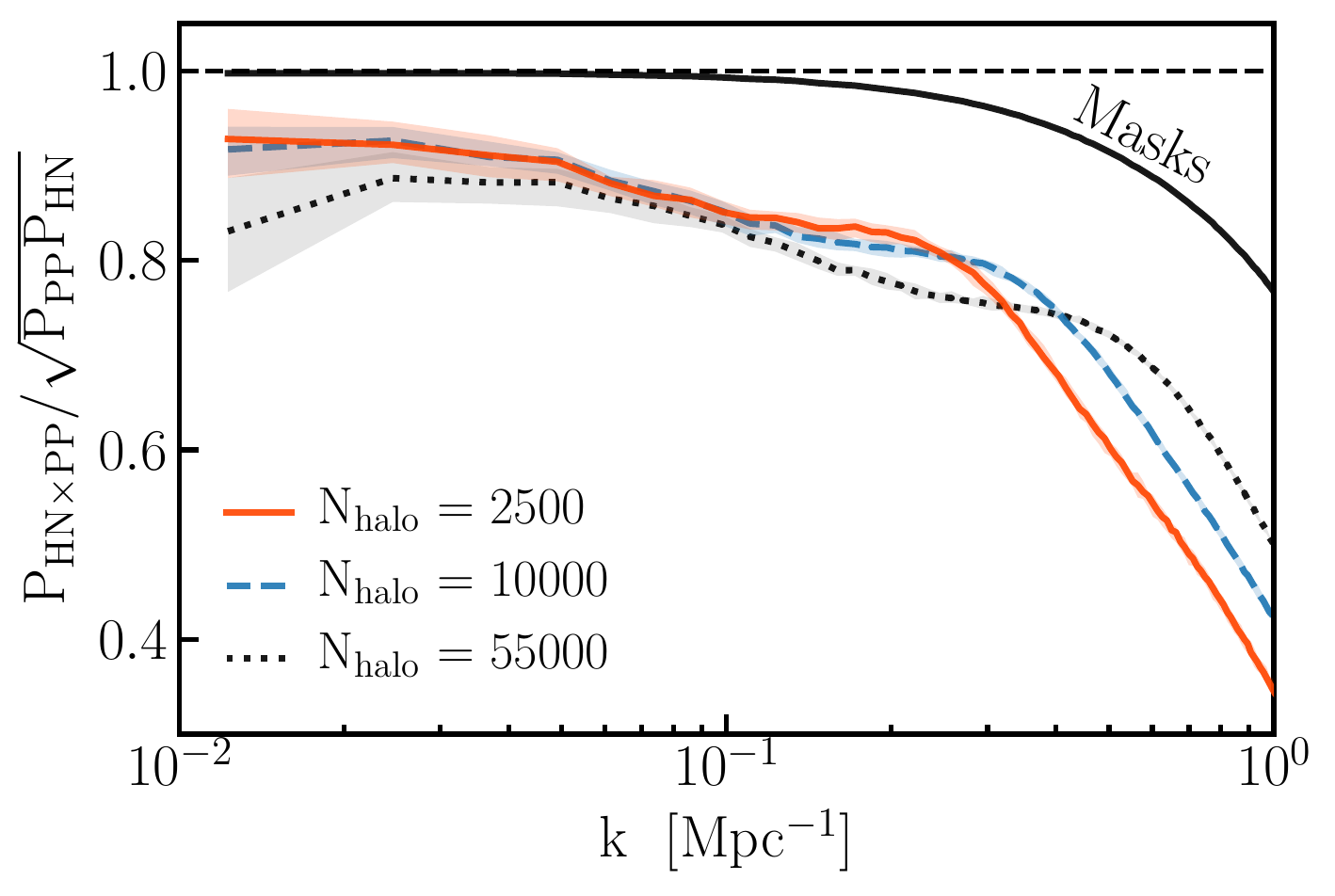}
\end{center}
\vspace{-0.3cm}
\caption{Halo power spectrum (Eq.~\ref{eq:powerspectrum}) of the 32 test simulations. (top) The ratio of the HaloNet power spectrum to that of the true haloes, for three number cuts. (bottom) The cross correlation coefficient (Eq.~\ref{eq:crosscorrelation}) for the same 3 number cuts. Also included is the cross correlation of the predicted and true masks, shown by the solid black line. Lines denote the mean, while the shaded areas represent the 1$\sigma$ uncertainty.}
\label{fig:halo_powerspectrum}
\end{figure}
%END FIGURE ----------------

Each simulation belonging to the test set has $\sim$60,000 true haloes and $\sim$55,000 HaloNet haloes above the minimum mass cut of 100 cells. Figure~\ref{fig:halo_massfunction} shows the abundance of HaloNet haloes compared to the true halo abundance, and compared to the \cite{tinker2008} and \cite{pressschechter} halo mass functions. We chose to compare to the \cite{tinker2008} mass function as it shares a similar definition of halo mass, measured by a spherical over-density calculation (SO) and not by friends-of-friends (FoF) as is common. We find that our results agree very well with the Peak Patch mass function across the entire mass range, and by extension also with Tinker. A key result is that HaloNet does not simply predict the same results of the \cite{pressschechter} mass function, which would be the case of a spherical collapse calculation using a simple density threshold, usually $\delta_c=1.686$. 

The over-prediction of mass for the largest HaloNet haloes comes from the fact that our halo finder can result in centre-of-mass positions slightly mis-centred towards overlapping haloes. As we used the true halo locations to measure the probability cutoff P$_{\rm cut}^{\rm shell}$, a mis-centered halo will now have increased probability when averaged in radial shells, as it now takes into account more voxels that belong to the neighbouring haloes. We observed that this over-prediction for the largest haloes also affected the neighbouring intermediate mass haloes, leading to a reduction of mass. 

To validate the Lagrangian halo finder (described in Section~\ref{sec:algo}), we tested by performing the halo finding measurements from the known centers of the true haloes. We found that the resulting catalogue was nearly identical visually and quantitatively to the original HaloNet catalogue, except for the highest mass haloes where we found a better fit to the true mass function, meaning that our simple prescription for finding haloes in the probability field works sufficiently well.

\subsection{Halo Clustering}
\label{sec:powerspectrum} 

Quantifying the clustering of haloes is done through calculating a spatial correlation function. The spatial correlation (or two-point) function is defined as the excess probability of finding a pair of haloes at a separation $\bf{r}$, when compared to what is expected for a random distribution. Instead of calculating the correlation function we used its Fourier transform, the halo power spectrum $P(\bf{k})$.

We calculate the halo power spectrum by binning haloes into a 512$^3$ grid (the same resolution as the initial density field) to create $\delta_h(\bf{x})$. The power spectrum is then defined as $\langle \delta_h({\bf{k}})\delta_h({\bf{k'}})\rangle = (2 \pi)^3 \delta_D^3({\bf{k}}+{\bf{k'}}) P_{hh}(k)$, which we can simply calculate for each linearly spaced bin $k_i$ through 

\begin{align}
\begin{split}
P(k_i) &= \int_{|{\bf{k}}| \in k_i} \frac{d^3{\bf{k}}}{V_{k_i}} \delta_h({\bf{k}})\delta_h(-{\bf{k}}) \\
&= \frac{1}{V_{cell}} \sum_{|{\bf{k}}| \in k_i} \frac{1}{n_{k_i}} \delta_h({\bf{k}})\delta_h(-{\bf{k}})
\label{eq:powerspectrum}
\end{split}
\end{align}
where the second equality holds when the calculation is performed on a discrete periodic grid such as we use in our analysis. $V_{cell}$ is the volume of a cell of the grid, and $V_{k_i} \approx 4\pi k_i^2 \Delta k$.

Figure~\ref{fig:halo_powerspectrum} shows the power spectrum and the cross correlation results for three number cuts, where the catalogues are first ranked in mass. The cross correlation is defined as 
\begin{equation}
r = \frac{P_{HN \times PP}}{\sqrt{P_{HN} \times P_{PP}}}
\label{eq:crosscorrelation}
\end{equation}
where HN denotes HaloNet, and PP denotes Peak Patch.

%BEGIN FIGURE -------------
\begin{figure*}
\begin{center}
\includegraphics[width=0.9\textwidth,trim={0 0 0 0},clip]{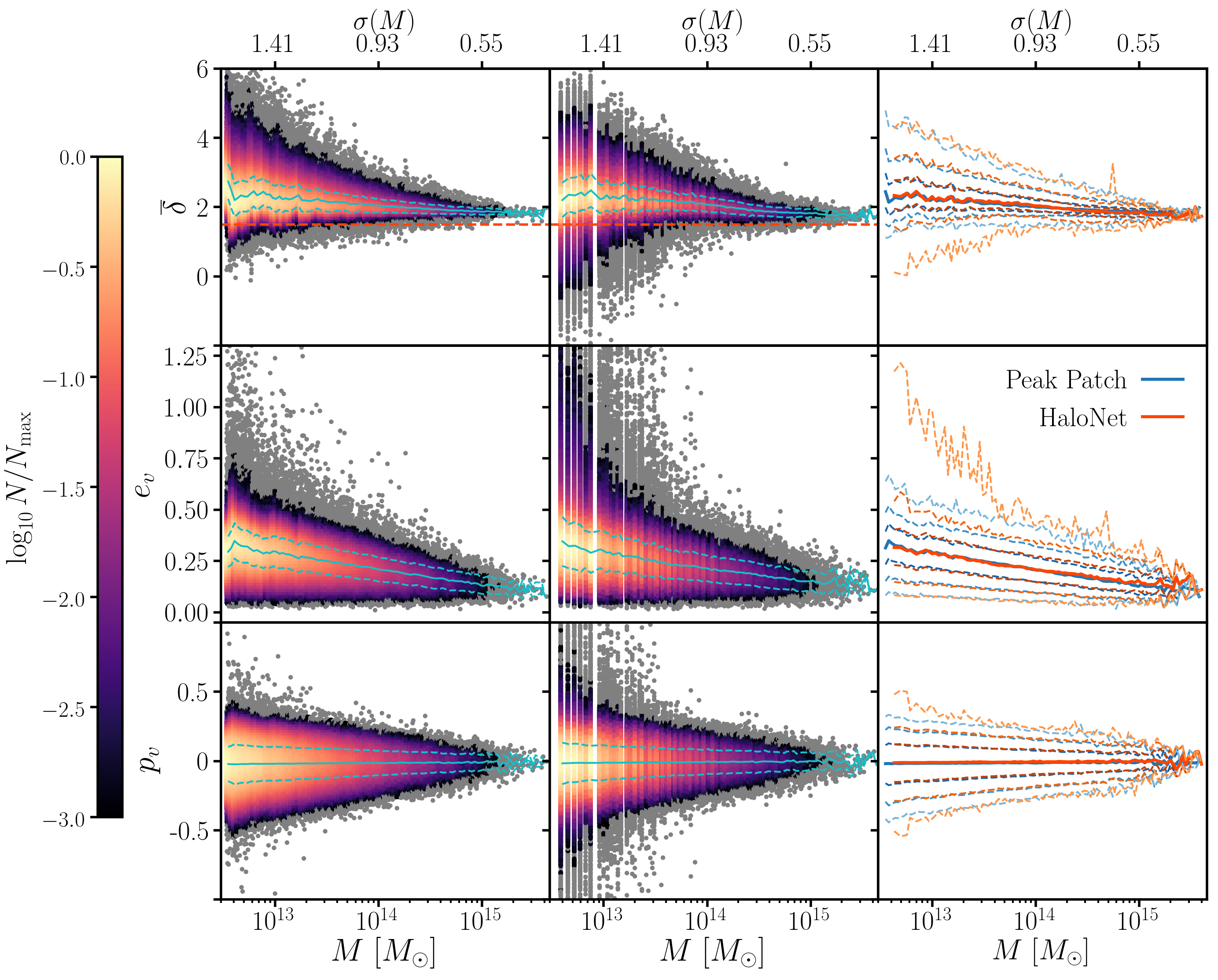}
\end{center}
\vspace{-0.3cm}
\caption{A comparison between the distributions $\bar{\delta}$, $e_v$, and $p_v$ as a function of halo mass, averaged over the 32 test simulations. The first and second columns show the distributions found in the Peak Patch and HaloNet simulations, respectively, while the third column overlays the median, and 1, 2, and 3 standard deviation regions from the median of the two methods. The colour in the first two columns gives the halo counts in bin normalized to the largest pixel in each panel, while regions below the range of the colour bar are greyed. The empty spaces between mass bins on the middle column are due to the discretization of small halo radii on a discrete grid in the HaloNet halo finder (see text below for details).}
\label{fig:halo_evpv}
\end{figure*}
%END FIGURE ----------------

We find that the power spectrum is within 5\% for the first number cut (most massive haloes) and when including all haloes. For all number cuts we find a roughly linear bias, meaning we reproduce well the shape of the power spectrum. We see that HaloNet performs well on large scales, including scales larger than the 64 Mpc sub-volumes that tile the full box.

\subsection{Halo Measurements}
\label{sec:measurements} 

Dark matter haloes provide a unique opportunity to study the predictions of Convolutional Neural Networks on an object-by-object basis. Based on the results of the previous sections, its clear that HaloNet is somehow searching for halo-like objects, but the degree to which it considers complicated features such as tidal forces is not. The Peak Patch algorithm explicitly includes tidal information by solving the ellipsoidal collapse equation (Eq. 2.21 of \cite{peakpatch}) to find the point of virialization for each halo at the target redshift. The essential inputs determining the collapse dynamics are the eigenvalues $\lambda_i, i=1\dots 3$, of the strain tensor,
\begin{equation}
s_{ij}(\mathbf{x}) = \frac{\nabla_i \nabla_j}{\nabla^2}\delta_m(\mathbf{x}),
\end{equation}
averaged within the radius of the peak. We can then compute the peak shear ellipticity, $e_v$ and prolaticity, $p_v$ as

\begin{align}
e_v &= \frac{1}{2\bar{\delta}_v}(\lambda_1 - \lambda_3),  
\\
p_v &= \frac{1}{2\bar{\delta}_v}(\lambda_1 - 2\lambda_2 + \lambda_3),
\end{align}
where $\bar{\delta}_v = \sum_i \lambda_i$ is the mean overdensity of the peak. The ordering $\lambda_1 \geq \lambda_2 \geq \lambda_3$ imposes the constraints $e_v \geq 0$ and $-e_v \leq p_v \leq e_v$ for each halo.

Figure \ref{fig:halo_evpv} show a comparison between the distributions $\bar{\delta}$, $e_v$, and $p_v$ as a function of halo mass, averaged over the 32 test simulations. We find that the HaloNet distributions closely resemble the Peak Patch over the entire range of this study, except for small deviations at the high and low mass ends. The empty spaces between mass bins on the middle column are due to defining HaloNet halos as spherical objects, with a mass equal to the number of cubic cells that are within their radius. Therefore, as the size of the sphere approaches the size of a cell, the discretization of the number of cells causes a noticeable separation in mass. For example, when going out in radius from a central cell on a 3D grid, one will first encounter 6 neighbours at a distance of 1R$_{\mathrm{cell}}$, then 8 more at a radius of $\sqrt{2}$R$_{\mathrm{cell}}$. This results in halo masses of 1M$_{\mathrm{cell}}$,7$_{\mathrm{cell}}$, and 15M$_{\mathrm{cell}}$, with none in between.

The low mass end represents the population where tides have the largest effect and so the properties deviate the most from the spherical collapse approximation (red line in Figure \ref{fig:halo_evpv}). On average, only haloes found in larger overdensities can collapse. While the predicted means and 1 standard deviation regions track the ground truth closely there, HaloNet occasionally selects more random regions, as evidenced by the tendency towards $\delta=0$ and the larger scatter in $e_v$, $p_v$. This suggests that HaloNet is to some degree aware of tidal forces but not to the accuracy of Peak Patch, at this stage of training. Furthermore, we have performed a matching analysis between the predicted and simulated catalogs, defining a match to have its centre inside and its size within 25$\%$ of the radius of its partner, which finds matches for 60$\%$ of Peak Patch haloes. While indeed the unmatched HaloNet haloes account for the scatter in Figure~\ref{fig:halo_evpv}, they maintain the same overall distributions. We note that the peak patch values of $\bar{\delta}$, $e_v$, and $p_v$, shown are not identical to the ones which dictate the homogeneous ellipsoidal collapse equations of Peak Patch, as we have recalculated them after the merging and exclusion steps of the algorithm, but they are overall very similar.

For the high mass end we see that $\bar{\delta}$ is sightly low with larger scatter, while $e_v$ is high (and noisy as well). This is most likely due to the effect of the 2 halo term on our halo finder, discussed in Section \ref{sec:algo}, biasing the more massive haloes larger. Alternatively, since the high mass end consists of the rarest events, to which the network is only exposed a handful of times, this could be the attempt of the network to extrapolate features learned on smaller spatial scales.

\section{Discussion \& Future Work}
\label{sec:discussion} 

In this work, we presented the first application of a volumetric deep Convolutional Neural Network (CNN) for simulation of dark matter halo catalogues. Our 5-level V-Net architecture is a powerful pixel-wise binary classifier for particles in the initial conditions (Section \ref{sec:train}), achieving a Dice coefficient of $\sim$0.93 in 72 hours of training. We showed how a simple geometric Lagrangian halo finder (Section \ref{sec:algo}) can be performed on the network output in order to create a dark matter halo catalogue that achieves high accuracy on the halo mass function, halo clustering, and individual halo properties, when compared to the true halo catalogue, across the entire mass range of the study. 

The effect of halo clustering on our halo finder has some noticeable effect on the accuracy, even after modifications helped to largely mitigate this (Section \ref{sec:algo}). A more expensive halo finder could increase the mass function accuracy, especially on the high mass end where Lagrangian regions belonging to halos have more overlap. Additionally, we find some evidence (Section \ref{sec:measurements}) that the network may be biased towards medium and small mass haloes. Re-weighting towards large mass haloes could help mitigate this \citep{gendiceloss}. However re-weighting schemes are computationally expensive, and furthermore disentangling the halo finder and biasing effects in highly clustered regions, on such a sparsely sampled population, will require the accuracy of N-body simulations as a training set.

We found that the HaloNet catalogue was consistent with the predictions of ellipsoidal collapse (Sections \ref{sec:massfunction}, \ref{sec:measurements}), the underlying principle in the ground truth Peak Patch simulations. Although we did observe increased scatter about this behaviour (Section \ref{sec:measurements}), we stopped the training of our network before observing a complete flattening of the loss function. Therefore increased training should result in higher accuracy, although at the point we stopped the learning rate is quite slow, so the required training is beyond the scope of this work.

This method allows for the simulation of large volumes of the universe for a very small computational cost and memory requirement. Although the computational cost is similar to that of the Peak Patch method that we used to construct out training set, it is orders of magnitude smaller than required by an N-body simulation. The high accuracy of the HaloNet prediction for Peak Patch gives motivation to instead train on N-body haloes traced back to the initial conditions, and it is this regime where the computational savings would be substantial. While the network architecture and algorithms presented here could be applied directly in that context, the training set is more expensive to compute. Furthermore, N-body haloes suffer from further complications such as complicated morphologies and initially disconnected regions ending up in the same halo. This suggests that exact N-body halo identification is a problem in semantic segmentation, where each halo is treated as a different class, intractable with a naive generalization of our method.

We have trained HaloNet at a single redshift and set of cosmological parameters. In its current form, therefore, our method suggests a process where the network is trained on a small sample of exact simulations (at a target redshift and cosmology) and is then used to generate large boxes and sets of independent realizations. However, the eventual goal would be to modify the network architecture to include fully-connected combinations of these parameters along side the CNN, trained at some grid sampling, to produce a cosmological realisation emulator.

\section*{Acknowledgments}

We thank Erik Spence for a great course on ``Advanced Neural Networks'' where we initially came up with the idea for this project and further developed our skills. We also thank Pranai Vasudev, Mohamad Ali-Dib, Charles Zhu, Kristen Menou, Dick Bond, Xin Wang, I-Sheng Yang, and Benjamin Wandelt for useful discussions along the way. Research in Canada is supported by NSERC. These calculations were performed on the GPC supercomputer at the SciNet HPC Consortium. SciNet is funded by: the Canada Foundation for Innovation under the auspices of Compute Canada; the Government of Ontario; Ontario Research Fund - Research Excellence; and the University of Toronto.

%\appendix

%\section{\label{app:appendix} Appendixes}

%We are now in Appendix~\ref{app:appendix}.

\bibliographystyle{mnras}
\bibliography{halonet}% Produces the bibliography via BibTeX.

% Don't change these lines
\bsp	% typesetting comment
\label{lastpage}
\end{document}